\documentstyle[twoside,fleqn,espcrc2,epsfig]{article}
\title{Pseudoscalar decay constant in heavy light systems}

\author{Joachim~Hein%
\address{Dept.~Physics \& Astronomy, University of Glasgow, G12 8QQ,
Scotland, UK; UKQCD Collaboration.}%
\thanks{In collaboration with A.~Ali~Khan, S.~Collins,
C.T.H.~Davies, C.~Morningstar, J.~Shigemitsu, J.~Sloan}}

\newcommand{\DD}{{\rm D}}

\newcommand{\imag}{{\rm i\hspace{0.13ex}}}

\newcommand{\bra}[1]{\langle #1 |}
\newcommand{\ket}[1]{| #1 \rangle}
\newcommand{\bgeq}{\begin{equation}}
\newcommand{\bgeqa}{\begin{eqnarray}}
\newcommand{\edeq}{\end{equation}}
\newcommand{\edeqa}{\end{eqnarray}}

\newcommand{\ainv}{a^{-1}}

\newcommand{\fps}{f_{\rm PS}}
\newcommand{\Gamps}{\gamma_5\gamma_0}
\newcommand{\lqcd}{\Lambda_{\rm QCD}}
\newcommand{\figwidth}{6.85cm}
\begin{document}
\begin{abstract}
We discuss the size of the higher order terms in the NRQCD
expansion of the pseudoscalar decay constant. Power law divergences
in the matrix elements contributing to $\fps$ are also investigated.
\end{abstract}
\maketitle

\section{Introduction}
When investigating the physics of the heavy light mesonic system within
the frame work of non-relativistic QCD (NRQCD), the action and observables are
expanded in powers of $\lqcd/M_Q$, $M_Q$ denoting the heavy
quark mass. 
If the regularisation, in our case the lattice cutoff, would be
removed, divergences will arise in the 
matrix elements and coefficients of the expansion.
To circumvent this, we keep the cutoff finite and use
an improved action to minimise residual effects on physical results.
Large residual divergences would become obvious as scaling violations
of the contributing matrix elements.

We use 278 quenched $12^3\times 24$ configurations with $\beta = 5.7$,
generously
provided by UKQCD. For the  light quarks the
tadpole improved clover action with $\kappa = 0.1400$ has been
used. This is slightly lighter than the strange quark mass as
determined from the $K$ meson, $\kappa_s|_K = 0.1398$ \cite{Prowland}. At 
$\beta = 6.0$ \cite{arifa} such a mismatch leads to a shift of 1\% in
$\fps$ which is negligible compared to other errors in our calculation. 
From $m_\rho$ one gets $\ainv = 1.103(11)(50)$~GeV \cite{hugh}.
For the heavy quarks we used a tadpole improved 
action corrected up to order ${\cal O}(1/M^2)$ in the bare heavy quark
mass, details on which can be found in \cite{edproc}. We use heavy
quark masses in the range $20 \ge aM \ge 0.6$, which ranges from heavy
quarks much heavier than the $b$ and slightly lighter than the $c$ quark.

All results presented are preliminary.

\section{Size of $\fps$ contributions}
The pseudoscalar decay constant $\fps$ is defined by the
matrix element
\bgeq
p_\mu \fps := \langle 0 | A_\mu|{\rm PS}\rangle\,,
\edeq
of a pseudoscalar meson state $\ket{\rm PS}$ and the axial vector current
$A_\mu$. In lattice NRQCD $A_\mu$ has to be matched by a set of local
operators $J_{A,\rm lat}^{(i)}$
\bgeqa \label{Aexpan}
A_0 &=& \sum_{i=0}^2 c_i(\alpha_s,aM) J_{A,\rm lat}^{(i)}\nonumber \\
&+& \sum_{i=3}^5 J_{A,\rm lat}^{(i)} + 
{\cal O}(\alpha_s^2,a^2,{\textstyle
\frac{\alpha_s}{M^2},\frac{a\alpha_s}{M}})\,,\\[1ex]
J_{A,\rm lat}^{(0)} &=& \bar q \Gamps  Q\,,\nonumber \\
J_{A,\rm lat}^{(1)} &=& -\frac{1}{2M} \bar q \Gamps (\vec \gamma \vec \DD Q)\,,\nonumber \\
J_{A,\rm lat}^{(2)}&=& \frac{1}{2M}(\vec \DD \bar q \vec \gamma)\Gamps Q\,,\nonumber \\
J_{A,\rm lat}^{(3)}&=& \frac{1}{8M^2}\bar q \Gamps \DD^2 Q\,,\nonumber \\
J_{A,\rm lat}^{(4)}&=& \frac{g}{8M^2}\bar q \Gamps\vec \sigma \vec B Q \,,\nonumber \\
J_{A,\rm lat}^{(5)}&=& -\frac{\imag g}{4M^2}\bar q \Gamps \gamma_0 \vec \gamma \vec E Q\,.\edeqa
$M$ denotes the bare heavy quark mass.
The $c_i$ are determined in one loop PT \cite{colinjunko}. Since 
the scale $q^*$ of the coupling constant $\alpha_s$ 
has not been calculated, we will average the final results over $aq^*
= 1$ and $\pi$. The difference is treated as a
systematic error of the pert.\ expansion.

\begin{figure}[tb]
\epsfig{file=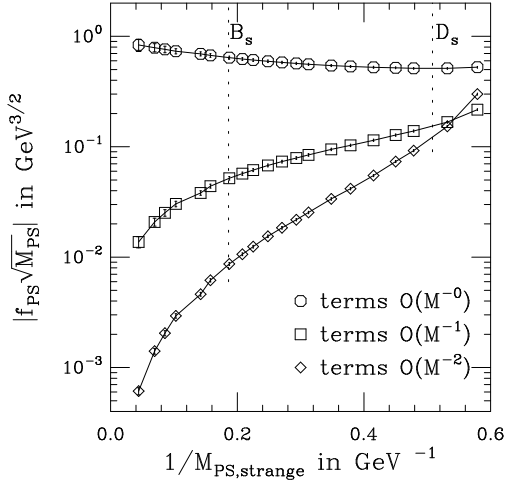, width=\figwidth}
\vspace*{-5mm}
\caption{\label{sizefig}Size of the tree level contribution to the
pseudoscalar decay constant versus strange pseudoscalar meson mass.}
\end{figure}

In figure~\ref{sizefig} we display the size of the matrix element
contributing to
$\fps$ in different orders of the bare heavy quark mass.
At the $b$-quark the different orders in $1/M$ are nicely suppressed
by one order of magnitude with respect to the previous. However
at the $c$-quark the ${\cal O}(\frac{1}{M^2})$ contribution has the
same size as the ${\cal O}(\frac{1}{M})$ part. 

If you look it in detail, this is not that surprising.
For the term in ${\cal O}(\frac{1}{M})$ we
observe a suppression 
of 0.36~GeV/$M$ at the $B_s$ and 0.29~GeV/$M$ at
the $D_s$ meson with respect to the leading term. 
For the ${\cal O}(\frac{1}{M^2})$ term the suppression
is 0.26~GeV$^2/M^2$ resp.\ 0.22~GeV$^2/M^2$. These numbers are
compatible with the expectation of a factor $\approx \lqcd/M$ for each 
order in $\frac{1}{M}$, but for the ${\cal O}(\frac{1}{M^2})$ 
term there is apparently a prefactor of $\approx 2$ to $3$.
At the $c$ quark $aM \le 1$, which could have some affect on the outcome,
requiring further investigation.

\section{Power law contributions}
\begin{figure}[tb]
\epsfig{file=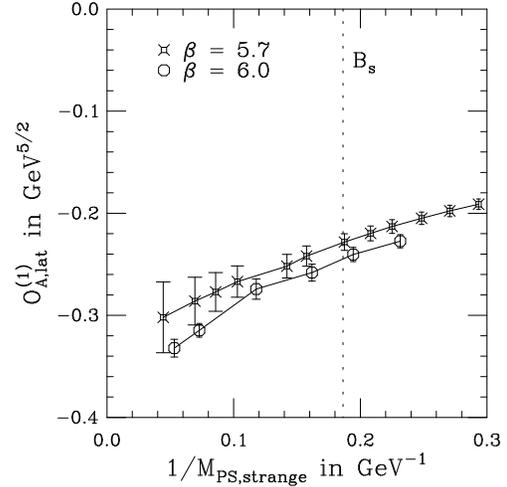, width=\figwidth}
\vspace*{-5mm}
\caption{\label{O1fig}
$O^{(1)}_{A,\rm lat}$ vs strange pseudoscalar meson mass.
}
\end{figure}

Due to operator mixing, the coefficients $c_i$ of eqn.~(\ref{Aexpan})
would diverge for $a \to 0$
as ${\cal O}(\frac{\alpha_s}{aM})$, ${\cal O}(\frac{\alpha_s}{(aM)^2})$,
\dots \cite{maiani}, which is unphysical. Therefore the divergences have to
cancel against similar terms arising from the matrix elements of the
$J^{(i)}_{A,\rm lat}$. To investigate the size of the unphysical part
of the matrix elements in comparison to their physical, we compare to the
results of \cite{arifa}, which were obtained at $a^{-1}(m_\rho) = 1.92$~GeV.
For the comparison we study
\bgeqa
O_{A,\rm lat}^{(1)} &=& \bra{0}M J^{(1)}_{A,\rm lat}\ket{\rm PS}\,,\nonumber\\
O_{A,\rm lat}^{(M2)}  &=& \bra{0}M^2 (J^{(3)}_{A,\rm lat}
\! +\! J^{(4)}_{A,\rm lat}\! +\! J^{(5)}_{A,\rm lat})\ket{\rm PS},
\edeqa
which have a reduced $M$ dependence and a non-zero
static limit. The result for $O_{A,\rm lat}^{(1)}$ 
is shown in figure~\ref{O1fig}. 
No scaling
violation is found within the quite small error bars. For the
ratio of these values we obtain 1.06(5) at the $B_s$, which seems
incompatible with a pure $\alpha_s/aM$ power law behaviour. 
Using $\alpha_s(aq^*=2)$ this would lead to a ratio of 1.39.
Our ratio of the $O_{A,\rm lat}^{(1)}$ is
independent of the quark mass and its error stays below 0.1 even for
the heaviest quark mass.

\begin{figure}[tb]
\epsfig{file=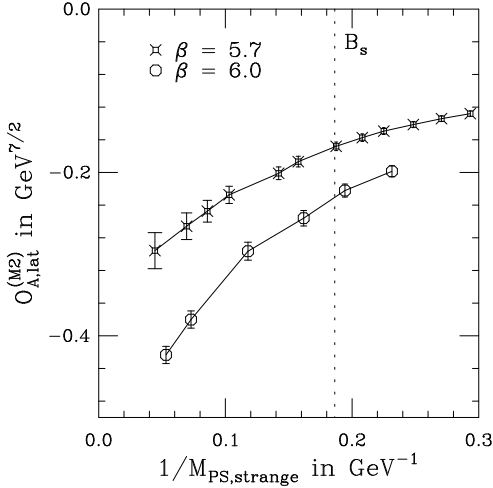, width=\figwidth}
\vspace*{-5mm}
\caption{\label{OM2fig}
$O^{(M2)}_{A,\rm lat}$ vs strange pseudoscalar meson mass.
}
\end{figure}

For $O_{A,\rm lat}^{(M2)}$ we observe some scaling violation as
displayed in figure~\ref{OM2fig}. At the $B_s$ the result 
on the fine lattice is by a factor of 1.35(7) larger
 than the one from the coarse one. However a power law
behaviour $\alpha_s/(aM)^2$ would lead to a factor of 2.4.
Again the scaling violations are independent of the heavy quark mass. 

\section{Pseudoscalar decay constant}
\begin{figure}[tb]
\epsfig{file=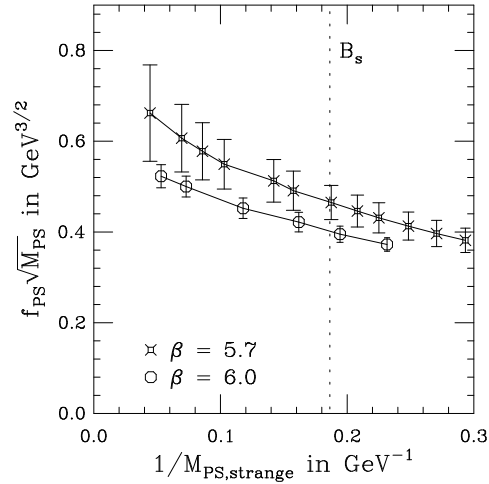, width=\figwidth}
\vspace*{-5mm}
\caption{\label{fratfig}
Strange pseudoscalar decay constants for two different
lattice spacings.}
\end{figure}

In fig.~\ref{fratfig} we show the scaling behaviour of  $\fps$ itself.
The error bars encompass the statistical errors and the uncertainty of
$q^*$ for both graphs, since the latter need not to be equal when
changing the $a$ value.
Apart from the heaviest meson masses, the error
bars on the $\beta = 5.7$ graph are dominated by the uncertainty of $q^*$.
The central values differ by $15\%\pm 10\%$, which is compatible with the 
expected ${\cal O}((a\lqcd)^2)$ discretisation correction.
Furthermore, a plot of the ratio of these results is
flat within error bars.

For the decay constant of the $B_s$ we obtain 
from the data set at $\beta=5.7$
\bgeq
f_{B_s} = 201(6)(15)(7){\rm \ MeV}\,.
\edeq
Here we only give the statistical error, the uncertainty
arising from $q^*$ and the determination of the bare $b$ quark
mass.

\section{Conclusion}
In this talk we discuss the size of the contributions to $\fps$ in a
large range of heavy quark masses, encompassing the $b$ and the $c$
quark. At the $b$ we find higher orders to be nicely  suppressed.
Scaling studies find no evidence for large power divergent contributions to 
higher order matrix elements, which appear to be dominated by their
physical ${\cal O}(\lqcd/M)$ resp.\ ${\cal O}(\lqcd^2/M^2)$ term.
For the decay constant itself we observe reasonable scaling behaviour.

A Marie Curie research fellowship 
by the European commission under ERB FMB ICT 961729,
and financial support of the collaboration by
NATO under CRG 941259 and by the US DOE are gratefully acknowledged.
The simulations were performed at NERSC.

\end{document}